\documentclass[usenatbib]{mn2e}

\advance\voffset by -1cm

\def\bm#1{\mbox{\boldmath{$#1$}}}
\def\bms#1{\mbox{\boldmath{$\scriptstyle#1_1$}}}
\def\i{\bm{i}} \def\j{\bm{j}} \def\k{\bm{k}}
\def\A{\bm{A}} \def\B{\bm{B}}
\def\p{\bm{p}} \def\q{\bm{q}}
\def\P{\bm{P}} \def\Q{\bm{Q}}
\def\R{\bm{R}} \def\r{\bm{r}} \def\n{\bm{n}}
 \def\F{\bm{F}}
\def\tr#1{\mathop{\rm re}\nolimits\left[#1\right]}
\def\frac#1#2{\hbox{$#1\over#2$}}
\def\half{\frac12}
\def\nablaQ{\nabla\!_{\bms Q}} \def\nablaP{\nabla\!_{\bms P}}

\title[The Kustaanheimo-Stiefel transform]
{Interpreting the Kustaanheimo-Stiefel transform \\
in gravitational dynamics}

\author[P. Saha]{Prasenjit Saha\\
Institute for Theoretical Physics, University of Z\"urich, \\
Winterthurerstrasse 190, 8057 Z\"urich, Switzerland}

\begin{document}

\maketitle

\begin{abstract}
  The Kustaanheimo-Stiefel transform turns a gravitational two-body
  problem into a harmonic oscillator, by going to four dimensions.  In
  addition to the mathematical-physics interest, the KS transform has
  proved very useful in $N$-body simulations, where it helps handle
  close encounters.  Yet the formalism remains somewhat arcane, with
  the role of the extra dimension being especially mysterious.  This
  paper shows how the basic transformation can be interpreted as a
  rotation in three dimensions.  For example, if we slew a telescope
  from zenith to a chosen star in one rotation, we can think of the
  rotation axis and angle as the KS transform of the star.  The
  non-uniqueness of the rotation axis encodes the extra dimension.
  This geometrical interpretation becomes evident on writing KS
  transforms in quaternion form, which also helps derive concise
  expressions for regularized equations of motion.
\end{abstract}

\begin{keywords}
celestial mechanics -- stellar dynamics
\end{keywords}

\section{Introduction}

The Kustaanheimo-Stiefel transform is a remarkable relation between
the two most important elementary problems in dynamics: under a
transformation of coordinates and time, a Kepler problem changes into
a harmonic oscillator. Especially noteworthy is that the collision
singularity in the Kepler problem is transformed into a regular point.
The name comes from the works by \cite{2001ASPC..237..279S} and
\cite{1965KS}, while the book by \cite{1971QB351.S758.....}, which is
largely devoted to the KS transform and its consequences, is perhaps
the best known source.  For a very short summary, see `regularization'
in \cite{2008gady.book.....B}.  An important application of the KS
transformation is in numerical orbit integration, where the
singularity-removal is used to great advantage for simulating dense
stellar systems with near collisions
\citep{1974CeMec..10..185A,1974CeMec..10..516A,1990CeMDA..47..375M,1993CeMDA..57..439M,1989ApJS...71..871J}.
Some recent papers also re-examine the formalism itself
\citep{2003JPhA...36.6963B,2006CeMDA..95..201W}.

In two dimensions there is a much simpler version of the KS transform
going back to \cite{1920LC}.  In the Levi-Civita transform, the
coordinate plane is read as the complex plane, and the complex square
root of the coordinate becomes the transformed coordinate.  The
geometrical interpretation is clear: the complex phase gets halved.
The KS transform is also a kind of square root, but in four
dimensions.  One wonders how the geometrical interpretation
generalizes.

It turns out that slewing a telescope is a convenient geometrical
analogy.  Suppose the telescope is at zenith, and we want to slew it
to a particular star in one rotation.  Normally we would simply move
along a great circle from the zenith to the star.  But we might prefer
a different rotation (to avoid crossing the moon, say).  For example,
we could choose the midpoint on the above great circle and rotate
about it by $180^\circ$.  In any case, the chosen rotation axis and
the rotation angle are effectively the KS transform of the star.  This
idea of rotation in three dimensions about a non-unique axis
generalizes the idea of halving the phase in a complex square root.

This paper attempts to provide some new insight into the KS transform
by providing some reformulations and new derivations of known results,
and especially to make the geometrical interpretation evident.

\section{Quaternions and Rotation}

Before considering KS theory, it is useful to review a concise
algebraic way of specifying rotations in three dimensions, not often
used in astrophysics but standard in computer graphics: quaternions.

Quaternions are a generalization of complex numbers.  The $\sqrt{-1}$
of complex numbers is replaced by three unit quaternions $\i,\j,\k$,
such that
\begin{equation}
\i\i = \j\j = \k\k = -1, \qquad \i\j\k = -1 .
\label{quatdef}
\end{equation}
From (\ref{quatdef}) it follows that $\i\j=\k=-\j\i$ and so on.  In
other words, quaternions are like a combination of dot and cross
products in vector algebra.  (Although historically quaternions came
before, having been invented by none other than W.R.~Hamilton of
Hamilton's equations.)

A general quaternion has the form
\begin{equation}
     \A = A_0 + A_x\i + A_y\j + A_z\k
\end{equation}
where we will call $A_0$ the real part.  A quaternion with no real
part is effectively a vector in three dimensions.

In analogy with complex numbers, we will use the following notation
for quaternion conjugates and absolute values.
\begin{eqnarray}
\tr{\A} &\equiv& A_0  \nonumber \\
\A^*    &\equiv& A_0 - A_1\i - A_2\j - A_3\k \\
A^2     &\equiv& \A^*\A = A_0^2+A_1^2+A_2^2+A_3^2 \nonumber
\end{eqnarray}
It is easy to see that $\tr{\A^*}=\tr{\A}$ and $(\A\B)^*=\B^*\A^*$, and
as a result
\begin{equation}
\tr{\A\B} = \tr{\B^*\A^*} = \tr{\B\A} .
\label{trperm}
\end{equation}

Rotation in quaternion notation is beautifully concise.  Say we want
to rotate a vector $\r$ by angle $\omega$ about a unit vector $\n$.
Using quaternion algebra the rotation is simply
\begin{equation}
\R^* \r \R
\label{qrot}
\end{equation}
where
\begin{equation}
\R = \cos\half\omega + \sin\half\omega\,\n .
\label{qrotang}
\end{equation}
Unlike the equivalent expression using Euler angles, the
expression (\ref{qrot}) has no coordinate singularities (or ``gimbal
lock'') and as a result is numerically more stable, which explains its
popularity in computer graphics.

For an arbitrary (i.e., non-unit) quaternion $\R$, the expression
(\ref{qrot}) amounts to a rotation combined with scalar multiplication.

It is possible to represent quaternions as matrices (though not
necessary, even for numerical work).  A familiar representation is in
terms of Pauli matrices
\begin{equation}
     \i=i\sigma_1  \quad  \j=-i\sigma_2  \quad  \k=i\sigma_3
\end{equation}
or
\begin{equation}
   \i = \left(\begin{array}{rr} 0 & i  \\ i &  0 \end{array}\right)  \quad
   \j = \left(\begin{array}{rr} 0 & -1 \\ 1 &  0 \end{array}\right)  \quad
   \k = \left(\begin{array}{rr} i & 0  \\ 0 & -i \end{array}\right)
\end{equation}
Pauli matrices are most important as operators on quantum two-state
systems (being Hermitian, whereas quaternions are anti-Hermitian). In
recent years the most exciting two-state quantum systems have been
Qbits in quantum computing.  It turns out that expressions of the type
(\ref{qrot}) appear in the description of quantum-computing gates
\citep[see][ who also provides a derivation of essentially the above
three-dimensional rotation formula]{2007qucosc.book}.

\section{The Kustaanheimo-Stiefel transform}

Let
\begin{equation}
\q = x\i + y\j + z\k
\end{equation}
denote a point in space.  The KS transform of $\q$ is the quaternion
\begin{equation}
\Q = Q_0 + Q_x\i + Q_y\j + Q_z\k
\end{equation}
the transformation formula being
\begin{equation}
\q = \Q^*\k\,\Q .
\label{Qtoq}
\end{equation}
A solution for $\Q$ is
\begin{equation}
\Q^{I} = {x\i + y\j + Z\k\over\sqrt{2Z}} \qquad
Z \equiv z + \sqrt{x^2+y^2+z^2}
\label{Qsol}
\end{equation}
as is easily verified by multiplication, following the quaternion
rules.  But $\Q^{I}$ is not unique, because changing to
\begin{equation}
\Q = (\cos\psi-\sin\psi\,\k) \, \Q^{I}
\label{gauge}
\end{equation}
leaves Equation~(\ref{Qtoq}) invariant.  Thus $\psi$ behaves like a gauge.

Everything so far is already in the literature.  The new result in
this paper is that we can readily visualize $\Q$, including its
non-uniqueness.

Comparing (\ref{Qtoq}) and (\ref{qrot}), it is evident that
$\Q$ is a rotator that takes the $z$ axis to $\q$.  To visualize $\Q$,
let us rewrite $\q$ as
\begin{equation}
\q = r ( \sin\theta\cos\phi\,\i  + \sin\theta\sin\phi\,\j + \cos\theta\,\k )
\end{equation}
where $r,\theta,\phi$ are the usual polar coordinates.  Rewriting $\Q^{I}$
in the solution (\ref{Qsol}) and simplifying, we have
\begin{equation}
\Q^{I} = \sqrt r ( \sin\half\theta\cos\phi\,\i  + \sin\half\theta\sin\phi\,\j +
                   \cos\half\theta\,\k ) .
\end{equation}
In other words, the zenith distance of $\Q^{I}$ is halfway along the great
circle from $\k$ to $\q$.  From (\ref{qrotang}) we see the rotation
angle $\omega$ would be $\pi$.  Now let us apply the gauge
transformation (\ref{gauge}) with $\psi=\pi/2$ to $Q^I$.  This gives
\begin{equation}
\Q^{II} = \sqrt r ( \cos\half\theta + \sin\half\theta\sin\phi\,\i
                    - \sin\half\theta\cos\phi\,\j )
\end{equation}
Now the implied rotation is by $\theta$, about an axis perpendicular
to both $\k$ and $\q$.  In general, we can write
\begin{equation}
\Q = \cos\psi\,\Q^I - \sin\psi\,\,\Q^{II}
\end{equation}
which is to say, $\Q$ could be anywhere on the great circle joining
$\Q^{I}$ and $\Q^{II}$.  The telescope-slewing analogy given above is
simply a description of the preceding three formulas.

An interesting special case is $\phi=0$, which gives
$\q=r(\cos\theta\,\k+\sin\theta\,\i)$ and $\Q^{I} = \sqrt r
(\cos\half\theta\,\k + \sin\half\theta\,\i)$.  Then $\Q^{I}$ is
effectively the complex square root of $\q$ (we need to read $\k$ as
the real axis and $\i$ as the imaginary axis).  In other words, the
planar case can be reduced to the Levi-Civita transform by a
suitable gauge.

Quaternion formulations of the KS transform have been discussed in
several authors: \cite{1971QB351.S758.....} mention quaternions but
appear to dislike them, while later authors \cite[for
example][]{1994CeMDA..60..291V,2006CeMDA..95..201W} are more
favourable.  The precise definition adopted for the transform varies,
but is equivalent to Eq.~(\ref{Qtoq}).  That $\Q$ represents a
rotation and shrinking/stretching of $\q$ is also known.
\cite{2003JPhA...36.6963B} specifically notes that the rotation axis
is unique in two dimensions but not in three.  But the explicit
description of the implied rotations, as above, appears to be new.

\section{The canonical momentum}

So far we have just discussed geometry, but of course the real
significance of the KS transform is dynamics, which we now
consider. Let
\begin{equation}
\p = p_x\i + p_y\j + p_z\k
\end{equation}
be the canonical momentum conjugate to $\q$.  We seek
\begin{equation}
\P = P_0 +P_x\i + P_y\j + P_z\k
\end{equation}
that will be canonically conjugate to $\Q$. Let us write
\begin{equation}
\tr{\p^* d\q} = \tr{\p^* \,d\Q^*\,\k\Q}+\tr{\p^* \Q^*\k \,d\Q}
\end{equation}
Using the identity (\ref{trperm}) we can rewrite the middle term as
$\tr{\p\,(d\Q^*\,\k\,\Q)^*}$. Since $\p=-\p^*$ and
$(d\Q^*\,\k\,\Q)^*=\Q^*\,(-\k)\,d\Q$ the term becomes becomes
$\tr{\p^*\Q^*\k\,d\Q}$.  Thus we have
\begin{equation}
\tr{\p^* d\q} = 2\tr{\p^* \Q^*\k \,d\Q}
\end{equation}
Now if we define
\begin{equation}
\P = -2\k\Q\p
\label{Pdef}
\end{equation}
we have
\begin{equation}
\tr{\p^* d\q} = \tr{\P^* d\Q}
\end{equation}
which is to say, $\p\cdot d\q=\P\cdot d\Q$.  Provided the Hamiltonian
depends on $\P,\Q$ only through $\p,\q$ and not on the gauge $\psi$,
the transformation $(\P,\Q)\rightarrow(\p,\q)$ is canonical.

To get an explicit expression for $\p$, we multiply
(\ref{Pdef}) on the left by $\Q^*\k$, obtaining
\begin{equation}
\p = {\Q^*\k \P \over 2Q^2} .
\end{equation}
Note that while we have to be careful about the order of
multiplication when $\i,\j,\k$ are involved, real numbers like $Q^2$
commute with everything.  Since $\p$ has no real part,
$\tr{\Q^*\k\P}=0$ identically.  We can think of it as a formal
constant of motion resulting from invariance with respect to $\psi$.

That $\P$ (as defined in Equation \ref{Pdef}, or equivalently)
completes a canonical transformation is a standard part of KS theory,
but the derivation of the canonical condition using quaternion
identities appears to be new.

\section{The two-body problem and the harmonic oscillator}

Let us now write the Kepler Hamiltonian
\begin{equation}
H = \half p^2 - 1/q
\end{equation}
in terms of KS variables.  Multiplying each of (\ref{Qtoq}) and
(\ref{Pdef}) by its quaternion conjugate, we have
\begin{equation}
q^2 = Q^4, \qquad P^2 = 4p^2 Q^2
\label{scalarPQ}
\end{equation}
and substituting these gives
\begin{equation}
H = \frac18 P^2/Q^2 - 1/Q^2 .
\end{equation}
We now use a device known in Hamiltonian dynamics as a Poincar\'e time
transformation.  This involves introducing a fictitious time variable
$s$, whose relation to $t$ we choose to be
\begin{equation}
   dt = Q^2\,ds .
\label{kepeq}
\end{equation}
Since $Q^2$ is the radial distance in the Kepler problem,
(\ref{kepeq}) is in fact Kepler's equation, and $s$ is the
eccentric anomaly.  In the fictitious time variable $s$, the equations
of motion are given by a new Hamiltonian
\begin{equation}
   \Gamma = Q^2(H-E) = \frac18 P^2 - EQ^2 - 1 
\label{GammaH}
\end{equation}
with $E$ being the constant initial value of $H$.  The
time-transformed $\Gamma$ Hamiltonian is zero along a trajectory, but
its partial derivatives are not zero.

The Hamiltonian $\Gamma$ is remarkable indeed.  For $E<0$ (bound
orbits) it is a harmonic oscillator.  Since $\Q$ has four components,
$\Gamma$ is like a mass on an isotropic spring in four Euclidean
dimensions.  Thus the well-known fact that the bound Kepler problem
has a dynamical $O(4)$ symmetry.  For the unbound case, the symmetry
group is different: formally the Lorentz group, but with a physical
meaning completely different from special relativity.  And---perhaps
most importantly---Hamilton's equations for $\Gamma$ are well-behaved
even at $Q=0$ (a collision).  This is known as regularization and was
the original motivation for KS theory.

The effect of an external force $\F$ is simple.  From (\ref{Pdef}) it
follows immediately that $\F$ will add an extra contribution of
$-2\k\Q\F$ to $d\P/dt$, which amounts to a contribution of
$-2Q^2\k\Q\F$ to $d\P/ds$.  Provided the external force is
non-singular, the equations of motion in $s$ remain regular.

\section{Regularizing the three-body problem}

Application of KS regularization to $N$-body simulations involve
expressing the gravitating system either as a tree-like hierarchy of
coupled two-body systems \citep{1989ApJS...71..871J} or as a chain
\citep{1990CeMDA..47..375M,1993CeMDA..57..439M}.  The basic idea can
be described using the three-body problem with all masses unity.  Here
again, quaternions enable a concise formulation.

In relative coordinates, the Hamiltonian for three unit gravitating
masses can be written \cite[cf. Eq. 12 in][]{1974CeMec..10..185A} as
\begin{equation}
    H = \frac12 p_1^2 + \frac12 p_2^2 + \p_1\cdot\p_2
      - {1\over q_1} - {1\over q_2}
\end{equation}
plus an additional potential $V(\q_1,\q_2)$.  Here $\q_1,\q_2$
expresses the position of the first and second body relative to the
zeroth body, while $\p_1,\p_2$ express the momenta of the first and
second bodies in the barycentric frame.  Meanwhile, $V(\q_1,\q_2)$
expresses the mutual interaction of the first and second bodies, plus
any external potential.  We can regard $V(\q_1,\q_2)$ as an external
potential, and since we already know how to deal with external forces,
we set $V$ aside and concentrate on $H$.

Now we introduce KS variables $\q_1=\Q_1^*\k\Q_1$ and so on.  Defining
\begin{equation}
   \Pi \equiv \tr{(\Q_1^*\k\P_1)^* \Q_2^*\k\P_2}
\label{dotasPi}
\end{equation}
we can write $\p_1\cdot\p_2$ as $\Pi/(4Q_1^2Q_2^2)$.  Applying a
Poincar\'e time transformation
\begin{equation}
   dt = Q_1^2 Q_2^2 ds \qquad \Gamma = Q_1^2 Q_2^2 (H - E)
\end{equation}
gives
\begin{equation}
    \Gamma = \frac18 P_1^2 Q_1^2 + \frac18 P_2^2 Q_2^2 + \frac14 \Pi
    - Q_2^2 - Q_1^2 - E Q_1^2 Q_2^2
\end{equation}
where $E$ is the value of $H$.  The $\Gamma$ Hamiltonian has no
denominators, and is thus regular for collisions with the zeroth body.
(We assume $V$ remains regular, that is to say, the first and second
bodies do not collide with each other or any other bodies than the
zeroth. In practice, simulations redefine the relative coordinates
whenever necessary, according to who is close to whom.)

For the equations of motion we need derivatives with respect to
quaternion components.  First we have $\nablaQ Q_1^2 = 2\Q_1$.
Slightly more subtle is $\nablaQ \tr{\Q_1^*\A}=\A$ if $\A$ is
independent of $\Q_1$.  Using this last identity, together with the
definition (\ref{dotasPi}) of $\Pi$, we derive
\begin{equation}
   \nablaP \Pi = -\k \Q_1 \Q_2^*\k\P_2 , \qquad
   \nablaQ \Pi = -\k \P_1 \P_2^*\k\Q_2 .
\label{noncomm}
\end{equation}
Hamilton's equations are then
\begin{eqnarray}
   {d\Q_1\over ds} &=& \frac14 Q_1^2\P_1 - \k \P_1 \P_2^*\k\Q_2 \nonumber\\
   {d\P_1\over ds} &=& \left( 2 + 2EQ_2^2 - \frac14 P_1^2 \right) \Q_1
                       + \k \Q_1 \Q_2^*\k\P_2
\end{eqnarray}
and similarly for $\P_2,\Q_2$.

\section{Discussion}

In dynamical astronomy the KS transformation is profound, but may
appear mysterious.  This paper attempts to make it less mysterious,
and hopefully therefore more useful, by explaining it in
three-dimensional geometric terms.  There are several possible
directions in which the KS transformation may turn out to be useful.

First, one can imagine new orbit integrators specialized to
nearly-Keplerian problems.  Work on dense stellar systems with near
collisions has already been mentioned \citep[for reviews see the
books][]{2003gnbs.book.....A,2003gmbp.book.....H}.  In the planetary
regime, which differs from the dense-stellar case in having few bodies
but many more orbital times, time transformations reminiscent of
(\ref{kepeq}) used for KS regularization have proved useful for highly
eccentric orbits \citep{1997CeMDA..67..145M,2002CeMDA..84..331E},
while some integration algorithms
\citep{1999CeMDA..74..287M,1999AJ....118.2532P} apply the time
transformation (\ref{kepeq}) implicitly.  Could the KS transformation
itself be exploited here?  \cite{2005AJ....129.2496F} has some further
ideas.

Second, it is conceivable that KS variables could simplify
perturbation theory.  Perturbation theory in classical celestial
mechanics \citep[see for example][]{2000ssd..book.....M} is
algebraically frighteningly complicated, basically because the natural
variables for the unperturbed and perturbed parts (being the Keplerian
action-angles and real-space coordinate) are related through an
implicit equation.  On the other hand, the action-angles of the
KS-transformed Kepler problem are explicitly related to space
coordinates---the implicit equation is transferred to the time
variable.  Could some major simplication be achieved through KS
variables?  Some progress has been made by \cite{2006NewA...11..366V}.

Third, the KS transformation might provide new insight into analogous
quantum problem.  \cite{RevModPhys.38.330,RevModPhys.38.346} derive
the symmetry groups of the bound and unbound Coulomb problem.  These
turn out to be the same four-dimensional symmetries as in KS theory.
Is the KS transformation implicit in that work?

\section{Acknowledgments}

I am grateful to thank Seppo Mikkola for introducing me to KS theory,
and to Marcel Zemp, Scott Tremaine, and the referee, J\"org Waldvogel,
for suggesting improvements in the manuscript.

\bibliographystyle{mn2e}
\bibliography{ms.bbl}

\end{document}